\documentclass[12pt, a4paper]{article}

\usepackage[utf8]{inputenc}
\usepackage[T1]{fontenc}
\usepackage{amsmath, amssymb, amsfonts, mathtools, mathrsfs}
\usepackage{graphicx}
\usepackage{caption}
\usepackage{subcaption}
\usepackage{float}
\usepackage{longtable, threeparttable, booktabs, multirow, tabularray}
\usepackage{textgreek}
\usepackage{rotating}
\usepackage{authblk}
\usepackage{booktabs,multirow,makecell}
\usepackage{siunitx}
\DeclareSIUnit\angstrom{\text{Å}}
\usepackage{titlesec}
\usepackage{hyperref}
\usepackage{xcolor}
\usepackage{enumitem}
\usepackage{csquotes}
\usepackage[font=small,labelfont=bf]{caption}
\usepackage[title]{appendix}
\usepackage[none]{hyphenat}
\usepackage{setspace}
\usepackage{afterpage, pdflscape, lscape}
\usepackage{algorithm}
\usepackage{siunitx}
\usepackage[version=4]{mhchem}
\usepackage{url}
\usepackage{algpseudocode}

\usepackage{amsmath}
\usepackage{booktabs}
\usepackage{tabularray}
\usepackage{xr-hyper}
\usepackage{hyperref}
\usepackage{cleveref}
\usepackage[
    backend=biber,
    style=nature,
    sorting=none,
    maxnames=10,
    doi=true,
    url=false
]{biblatex}

\usepackage{geometry}
\titleformat{\section}{\normalfont\Large\bfseries}{\thesection.}{1em}{}

\title{\begin{spacing}{1.1}
\textbf{Coupled Structural and Electronic Requirements in $\alpha$-FASnI$_3$ Imposed by the Sn(II) Lone Pair}
\end{spacing}}

\author[1]{Mridhula Venkatanarayanan}
\author[1]{Vladislav Slama}
\author[1]{Madhubanti Mukherjee}
\author[1]{Andrea Vezzosi}
\author[1]{Ursula Rothlisberger$^{*}$}
\author[1]{Virginia Carnevali$^{\dagger}$}

\affil[1]{\small Laboratory of Computational Chemistry and Biochemistry, 
Institute of Chemical Sciences and Engineering, 
Swiss Federal Institute of Technology (EPFL), Lausanne, Switzerland}

\affil[ ]{\small $^{*}$ursula.roethlisberger@epfl.ch}
\affil[ ]{\small $^{\dagger}$virginia.carnevali@epfl.ch}

\date{}
\begin{document}
\maketitle
\sloppy
\begin{abstract}
$\alpha$-Formamidinium-tin-iodide ($\alpha$-FASnI$_3$) is a leading candidate for lead-free photovoltaic applications, adopting a nearly cubic structure at room temperature, but its stability remains limited by oxidation-driven degradation. Reliable first-principles modelling of the photovoltaic $\alpha$-phase is further complicated by inconsistent structural models and levels of theory in the literature. Here, we identify the structural and electronic requirements needed for a physically sound description of $\alpha$-FASnI$_3$, whose behaviour is governed by a pseudo-Jahn-Teller (PJT) instability arising from the stereochemically active Sn(II) 5$s^2$ lone pair.

Using 0~K relaxations, cross-code hybrid-functional benchmarks, and finite-temperature ab initio molecular dynamics, we show that a 4$\times$4$\times$4 supercell with randomly oriented FA$^+$ cations (4$\times$4$\times$4$^*$) is the smallest model that removes macroscopic dipoles, preserves cubic symmetry, recovers local octahedral tilts, and captures the characteristic PJT-driven Sn off-centering. Accurate band edges and a reliable band gap require a PBE0-level hybrid functional with spin-orbit coupling to treat Sn relativistic effects, together with nonlocal dispersion (rVV10) to capture the enhanced Sn-I covalency. Finite-temperature simulations reveal that Sn off-centering remains local, $\langle 111\rangle$-oriented, and robust against thermal fluctuations, and that reproducing the experimental 300~K band gap requires a 6$\times$6$\times$6 supercell. These results define the essential ingredients for reliable modelling of $\alpha$-FASnI$_3$ and provide a rigorous foundation for studying lone-pair-driven physics in tin halide perovskites.

\end{abstract}

\section*{Introduction}

Organic-inorganic metal halide perovskites (MHPs) have achieved remarkable progress in photovoltaic performance, with power conversion efficiencies rising from 3.8\% to over 25\% in just over a decade~\cite{c1,c2,c3}. These ABX$_3$ materials, where A is an organic or alkali-metal cation, B a divalent metal cation, and X a halide anion, combine strong optical absorption and efficient charge transport~\cite{c1,c4}. While lead-based perovskites deliver the highest efficiencies~\cite{c5,c6,c7}, concerns over Pb toxicity have motivated the search for lead-free alternatives~\cite{c8,c9,c10,c11,c12}. Tin (Sn) has emerged as the most promising candidate due to its similar $ns^2$ configuration, which underpins defect tolerance and favorable carrier mobility. Sn-based perovskites exhibit a direct band gap of 1.2-1.4~eV~\cite{c14}, close to the Shockley-Queisser optimum of approximately 1.34~eV~\cite{c15}, making them ideal for single-junction photovoltaics. However, the intrinsic instability of Sn$^{2+}$, which readily oxidizes to Sn$^{4+}$ during fabrication and air exposure, produces vacancies and deep traps that degrade performance~\cite{c16_1}. Replacing the methylammonium (MA$^+$) cation with formamidinium (FA$^+$) partially mitigates these effects by relaxing octahedral tilts and reducing lattice strain, leading to higher carrier mobility and improved stability. As a result, the power conversion efficiencies of FA-Sn perovskites have increased from about 2\% in 2014 to nearly 13\% in 2021~\cite{c3}. Yet oxidation-induced defects persist, showing that FA substitution slows but does not eliminate the underlying degradation pathways.

Understanding why these degradation pathways persist requires probing the underlying atomistic mechanisms. In FASnI$_3$, Sn$^{2+}$ oxidation, defect formation, and additive-lattice interactions occur at buried interfaces, grain boundaries, and during crystallization. While SnF$_2$ is widely used to suppress oxidation~\cite{c17}, excess additive can segregate at grain boundaries and form secondary phases~\cite{dft_6}. Other halide additives, such as SnCl$_2$ and SnBr$_2$~\cite{c19,c20,c21,c22}, influence nucleation kinetics and lattice strain, yet the microscopic origins of these effects remain poorly understood. Morphological features, including pinholes, grain-size variation, and surface terminations, critically affect carrier lifetimes and ion migration, but direct experimental correlation with synthesis conditions is challenging. In this context, theoretical modeling plays a vital role in complementing experiments by providing atomistic-level insights that are otherwise inaccessible.

Computational studies of FASnI$_3$ and related compounds (MASnI$_3$, CsSnI$_3$) employ a range of approaches, including density functional theory (DFT) \cite{dft_1, dft_2, dft_3, dft_4, dft_5, dft_6}, time-dependent DFT (TDDFT) \cite{tddft_1, dft_6, tddft_2_namd}, nonadiabatic molecular dynamics (NAMD) \cite{namd, tddft_2_namd}, and ab~initio and classical molecular dynamics (MD) \cite{MD_1, MD_2, MD_3}. NAMD captures the influence of phonon scattering and defect migration on carrier mobility and nonradiative recombination, while TDDFT provides access to excited-state properties such as electronic state overlap, bandgap evolution, and recombination dynamics under A-site cation modification \cite{tddft_2_namd}. Within DFT, the nudged elastic band (NEB) method has been applied to determine formation and migration barriers of point defects, offering insight into stability under light and oxygen exposure \cite{NEB, tddft_2_namd}. MD simulations have clarified how processing conditions, such as growth and annealing temperatures, affect the structural, electronic, and defect characteristics of mixed-cation Sn iodide perovskites \cite{MD_3}. In the broader context of metal halide perovskites, high-throughput DFT combined with machine learning (ML) has been used to predict defect formation energies and charge transition levels, providing a data-driven framework for screening impurities and assessing their impact on optoelectronic performance \cite{ML, ML_2}. However, these ML approaches are only as reliable as the quantum-mechanical data used to train them, so high-quality first-principles studies of FASnI$_3$ are still required.

Compared with FAPbI$_3$, the computational literature on FASnI$_3$ remains limited and inconsistent, with substantial variation in structural models and levels of electronic theory. Although FASnI$_3$ is experimentally cubic in its stable black phase ($\alpha$-FASnI$_3$), some studies adopt tetragonal cells \cite{dft_6} by analogy with MASnI$_3$, which is experimentally tetragonal, while others use orthorhombic structures \cite{dft_7} or impose symmetry constraints that prevent relaxation into the cubic phase \cite{dft_1, dft_8}. System size further contributes to discrepancies: most calculations use the primitive or unit cell, enforcing ordered FA$^+$ orientations and suppressing the dynamic disorder observed experimentally \cite{dft_1, dft_4, dft_6}. Larger 2$\times$2$\times$2 supercells improve sampling but remain too small to capture long-range correlations \cite{dft_7, dft_8}. Methodological variation also arises from the choice of exchange-correlation functional. The generalized gradient approximation (GGA), typically in the Perdew-Burke-Ernzerhof (PBE) form \cite{pbe}, underestimates the band gap, and this error is amplified when spin-orbit coupling (SOC) is included to account for relativistic effects in the heavy tin atoms. Screened hybrids such as Heyd-Scuseria-Ernzerhof (HSE06) \cite{hse} mitigate this underestimation \cite{dft_1}, whereas global hybrids like PBE0 \cite{pbe0} have been used only for density-of-states alignment to experimental spectra \cite{pbe0_exp}, rather than for fully predictive band-structure calculations. Many-body approaches have likewise not yet been systematically applied to FASnI$_3$. Dispersion effects are also rarely examined in detail. Although empirical and semi-empirical corrections are sometimes included \cite{dft_6, dft_1}, nonlocal van der Waals treatments remain largely absent, even though the stronger covalency of the Sn-I bond suggests that long-range correlation may play a non-negligible role. As a result, reported band gaps and electronic structures vary widely, reflecting the combined influence of structural and methodological choices across the literature.

Insights from previous first-principles studies on the cubic photoactive phase of FAPbI$_3$ ($\alpha$-FAPbI$_3$) \cite{FAPbI3} are extended here to $\alpha$-FASnI$_3$, guiding the modeling strategy adopted in this work. The earlier results demonstrated that accurate first-principles simulations of hybrid halide perovskites require both structural realism and an appropriate level of electronic theory tailored to the material under study. Structurally, small simulation cells were found to constrain FA$^+$ cations into artificial ordered patterns, distorting the cubic lattice, inducing spurious dipoles, and suppressing long-wavelength octahedral tilt modes.These artifacts were eliminated only when large supercells, at least 4$\times$4$\times$4 (768 atoms), were used, with FA$^+$ cations oriented pseudo-randomly by assigning their C-H axes to one of the eight equivalent $\langle111\rangle$ directions. In such configurations, dipole contributions cancel, average cubic symmetry is restored, and FA$^+$ dynamics naturally converge the band gap. On the electronic side, the $\alpha$-FAPbI$_3$ study showed that the appropriate level of theory depends on system size and temperature. In Pb-based perovskites, SOC significantly narrows the band gap, but this effect is nearly compensated by the PBE0 overestimation at 0~K, yielding good agreement with experiment. At finite temperature, large-cell PBE simulations ($\geq$6$\times$6$\times$6) with full thermal sampling at 300~K reproduced the experimental band gap without hybrid functionals, indicating that accurate results can be achieved when structural and dynamical effects are properly represented.

Building on these insights, we extend the same framework to $\alpha$-FASnI$_3$, which is structurally similar to $\alpha$-FAPbI$_3$ but shows stronger Sn-I covalency and greater stereochemical activity of the Sn 5s$^2$ lone pair. Large supercells remain important to capture realistic FA$^+$ dynamics and remove finite-size artifacts. The enhanced bond covalency and slightly stiffer lattice suppress large-amplitude octahedral tilts, which may reduce the sensitivity of structural order to cell size, making it necessary to assess how large a model is required for converged orientational and electronic properties. On the electronic side, SOC is weaker than in Pb halides, so the near-cancellation between SOC-induced gap narrowing and PBE0 overestimation seen in $\alpha$-FAPbI$_3$ does not hold here. Identifying the most suitable level of theory, whether within GGA, hybrid, or meta-GGA frameworks combined with SOC, is essential to describe exchange, correlation, and relativistic effects accurately. The stronger Sn-I covalency and active lone pair also alter the balance between short-range bonding and long-range dispersion, making the reliability of empirical van der Waals corrections uncertain and motivating comparison with nonlocal correlation functionals that include dispersion more consistently.

\section*{Results}
\subsection*{Characterization of $\alpha$-FASnI$_3$ at 0~K}

\begin{table}[h!]
\centering
\scriptsize
\renewcommand{\arraystretch}{1.6}   
\setlength{\tabcolsep}{3pt}         

\resizebox{\linewidth}{!}{%
\begin{tabular}{
  l
  c c c c
  c c c c
}
\toprule
\multirow{2}{*}{Property}
  & \multicolumn{4}{c}{$2\times 2\times 2$}
  & \multicolumn{4}{c}{$4\times 4\times 4$} \\
\cmidrule(lr){2-5} \cmidrule(lr){6-9}
  & \rotatebox{0}{relax} 
  & \rotatebox{0}{vc-relax}
  & \rotatebox{0}{relax$^{*}$} 
  & \rotatebox{0}{vc-relax$^{*}$}
  & \rotatebox{0}{relax} 
  & \rotatebox{0}{vc-relax}
  & \rotatebox{0}{relax$^{*}$} 
  & \rotatebox{0}{vc-relax$^{*}$} \\
\midrule

\makecell{Sn \\ off-centering (\si{\angstrom})}
  & $0.20 \pm 0.00$ & $0.29 \pm 0.00$ & $0.22 \pm 0.04$ & $0.13 \pm 0.06$
  & $0.21 \pm 0.00$ & $0.26 \pm 0.00$ & $0.25 \pm 0.09$ & $0.24 \pm 0.08$ \\

\makecell{Off-centering \\ orientation order}
  & $1.00$ & $1.00$ & $0.85$ & $0.27$
  & $1.00$ & $1.00$ & $0.24$ & $0.10$ \\

\makecell{Sn–I–Sn \\ angle (\si{\degree})}
  & $171.83 \pm 2.17$ & $169.95 \pm 4.03$ & $168.15 \pm 7.10$ & $168.59 \pm 7.31$
  & $171.41 \pm 2.20$ & $170.77 \pm 3.74$ & $166.50 \pm 6.43$ & $166.29 \pm 7.03$ \\

\makecell{Octahedral \\ anisotropy index}
  & $0.06 \pm 0.00$ & $0.11 \pm 0.00$ & $0.08 \pm 0.02$ & $0.05 \pm 0.02$
  & $0.06 \pm 0.00$ & $0.10 \pm 0.00$ & $0.11 \pm 0.03$ & $0.10 \pm 0.02$ \\

\makecell{Normalized orbital \\ overlap proxy}
  & $0.97 \pm 0.01$ & $0.95 \pm 0.03$ & $0.92 \pm 0.08$ & $0.95 \pm 0.09$
  & $0.97 \pm 0.01$ & $0.96 \pm 0.03$ & $0.91 \pm 0.07$ & $0.91 \pm 0.08$ \\

\addlinespace[2pt]
\bottomrule
\end{tabular}%
} % end resizebox
\caption{Structural properties of $\alpha$-FASnI$_3$ obtained from geometry-optimized (relax) and variable-cell-relaxed (vc-relax) configurations for 2×2×2 and 4×4×4 supercells. Reported values correspond to the mean ± standard deviation averaged over all SnI$_6$ octahedra in the optimized structures. Configurations marked with an asterisk (*) contain randomly oriented FA molecules. Definitions of the listed parameters are provided in the Discussion section.}
\label{fasni3_structural}
\end{table}

To identify modeling choices appropriate for $\alpha$-FASnI$_3$, we carried out a series of 0~K simulations spanning system sizes from the primitive unit cell (12 atoms) to 6$\times$6$\times$6 supercells (2592 atoms). This range allows us to assess how finite-size effects influence lattice distortions and FA$^+$ orientational behavior; the 12-atom cell, although too small to accommodate octahedral tilts, is included as a reference because it remains widely used in the literature. Each system was optimized using either fixed-cell relaxations, which maintain cubic symmetry, or variable-cell relaxations, which allow the lattice to respond to the local bonding environment. We considered two representative FA$^+$ arrangements: an all-aligned (AA) configuration, where all dipoles point in the same direction, and a random configuration that distributes orientations uniformly between 0 and $2\pi$, thereby restoring overall cubic symmetry and cancelling the net dipole moment. PBE-optimized structures were then used for single-point electronic calculations including spin-orbit coupling (PBE+SOC) and hybrid exchange (PBE0, PBE0+SOC), with k-point meshes converged for each cell size and $\Gamma$-point sampling used for the largest systems. Dispersion interactions were included using Grimme’s D3 scheme \cite{D3} with zero damping. The resulting trends in band gaps and band-edge positions are summarized in Supplementary Tables~\ref{tab:fasni3_gaps} and~\ref{tab:fasni3_edges}.

Understanding which structural models are physically meaningful for $\alpha$-FASnI$_3$ requires considering the electronic origin of lattice distortions in Sn(II) halide perovskites. According to pseudo-Jahn-Teller (PJT) theory~\cite{PJTE_1}, $ns^{2}$ cations undergo off-centering when lattice distortions allow vibronic coupling between the filled $s^{2}$ ground state and low-lying $p$ states. The revised lone-pair (RLP) model~\cite{rlp} refines this by showing that the relevant ground-state orbital is the antibonding cation-$s$/anion-$p$ state at the valence edge, which mixes with cation-$p$ only after symmetry breaking. Together, PJT and RLP imply that in $\alpha$-FASnI$_3$ the strong Sn-5$s$/I-5$p$ antibonding and the availability of Sn-5$p$ states favour locally off-centred Sn environments, where symmetry breaking enables the asymmetric, stereochemically active 5$s^{2}$ lone pair to emerge.

Experiments on CsSnBr$_3$ show that even though the structure appears cubic on average, the local environment is strongly distorted: Sn$^{2+}$ undergoes large-amplitude dynamic off-centering with a clear preference for $\langle 111\rangle$, and these distortions weaken the Sn-Br $\sigma$-overlap, shifting the band gap to higher energy~\cite{PJTE_2}. Follow-up work shows that this behaviour is not unique to CsSnBr$_3$ but is a general feature of Sn(II) halide perovskites, which sit on a very flat, multidimensional energy landscape that accommodates several low-symmetry SnX$_6$ geometries with almost the same energy~\cite{PJTE_3}. Taken together, these results show that Sn off-centering is a built-in electronic instability of Sn$^{2+}$ rather than something driven by thermal motion or A-site dynamics. This provides the correct baseline for thinking about finite-size effects, dipolar fields, and artificial structural coherence when constructing supercell models and setting band-gap convergence criteria for $\alpha$-FASnI$_3$.

In the all-aligned FA simulation cells, the collective orientation of the FA molecules generates a macroscopic polarization that forces all Sn atoms to displace along the same $\langle 111\rangle$ direction (off-centering orientation order = 1), leading to coherent displacements that are field-driven rather than intrinsic (see Table~\ref{fasni3_structural}). Under cubic-constrained relaxations, we hypothesise that this built-in dipole field lowers the conduction-band minimum (CBM) through a Stark-like effect, producing an artificial band-gap narrowing to nearly half the experimental value. A similar trend was reported by Ji et~al.~\cite{electric_field}, who showed that applying an external field along $\langle 111\rangle$ in CH$_3$NH$_3$PbI$_3$ causes the CBM to shift downward while the valence-band maximum (VBM) is comparatively unaffected. When the lattice is allowed to relax in the vc-relax calculations, the structure distorts slightly to counteract the macroscopic FA dipole, as reflected in the increase in the octahedral anisotropy index. This partial screening of the internal field raises the CBM by approximately 0.2~eV, leading to a correspondingly larger band gap, while the VBM shifts upward by only about 0.05-0.10~eV across the cell sizes studied.

Random FA orientations give a clean way to isolate the intrinsic local distortions in $\alpha$-FASnI$_3$, because they remove the macroscopic polarization field that dominates in the all-aligned configurations. Each FA$^+$ still creates a local electric field that is asymmetric around the Sn site and breaks inversion symmetry, so small Sn off-centerings and local octahedral tilts still occur. However, since the FA dipoles point in different directions, these local fields vary from site to site. The resulting distortions are therefore incoherent: the Sn displacements and tilt angles fluctuate in direction and largely cancel out when averaged over a sufficiently large supercell. In this limit, the magnitude of the Sn off-centering becomes independent of system size and reflects the underlying pseudo-Jahn-Teller instability of the Sn$^{2+}$ lone pair, while the overall symmetry of the crystal remains cubic.

The 2$\times$2$\times$2* supercell does not reach this regime. Because it is too small, the local distortions repeat periodically and become artificially coherent, which leads to unphysical alignment of tilt patterns and Sn displacements. This coherence amplifies the structural distortions and produces large shifts of the band edges, but these shifts are finite-size artefacts rather than a consequence of the intrinsic PJT physics. In contrast, the 4$\times$4$\times$4* supercell is large enough for the distortions to remain uncorrelated. The FA dipoles cancel statistically, the band edges change only minimally upon vc-relaxation, and the Sn off-centering remains finite even though its direction varies randomly across the cell. This size-independent off-centering reflects the intrinsic, field-free PJT behaviour expected for Sn$^{2+}$ and is no longer masked by macroscopic dipoles or periodicity effects. For this reason, the 4$\times$4$\times$4* cell represents the minimum supercell size needed to recover the correct physical behaviour of $\alpha$-FASnI$_3$ at 0 K.

This interpretation is further supported by the evolution of the FA dipole moment and lattice metrics with system size, as summarized in Supplementary Table~\ref{mse_dip}. In the all-aligned systems, the net polarization decreases from 3.3~D per ABX$_3$ in the 2$\times$2$\times$2 cell to 0.9~D in the 4$\times$4$\times$4 cell, but the mean squared error (MSE) relative to the cubic reference increases, indicating that the lattice must distort more strongly to oppose the uniform FA field as the cell grows. In the random systems (2$\times$2$\times$2* and 4$\times$4$\times$4*), the net polarization decreases from 2.1~D to 0.36~D per ABX$_3$, and this cancellation of FA dipoles is accompanied by a steady reduction in MSE, showing that the average structure becomes progressively more cubic as the sampling of FA orientations improves.The 4×4×4 supercell is the smallest system in which FA dipoles cancel, cubic symmetry is recovered, and Sn off-centering becomes size-independent, indicating that finite-size artefacts have been removed. We therefore use 4×4×4* as the minimal reliable model for $\alpha$-FASnI$_3$.

\begin{table}[ht]
\centering
\resizebox{\textwidth}{!}{%
\begin{tabular}{llccccccc}
\toprule
Simulation cell & Dispersion scheme & PBE & PBE+SOC & PBE0 & PBE0+SOC & HSE06 & HSE06+SOC & k-grid \\
\midrule
4$\times$4$\times$4   & DFT-D3     & 0.97
& 0.70 & 2.01 & 1.70 & 1.41 & 1.21 & 1 \\
\midrule
\multirow{3}{*}{4$\times$4$\times$4$^{*}$} 
  & DFT-D3     & 0.81 & 0.49 & 1.84 & 1.48 & 1.25 & 0.90 & 1\\
  & DFT-D3(BJ) & 0.69 & 0.35& 1.69& 1.31 & 1.10 & 0.74 & 1 \\
  & DFT-D3$^{\dag}$      & 1.08 &(0.78)  & 2.07 & (1.77) & - & -& 1 \\
  & rvv10$^{\dag}$      & 0.85 & (0.55) & 1.76 & (1.46) & 1.33 & (1.03) & 1 \\
\bottomrule
\end{tabular}%
}
\caption{Kohn-Sham band gaps (eV) for FASnI$_3$ in the 4$\times$4$\times$4 (768-atom) supercell across different dispersion corrections and exchange-correlation functionals. 
Values are obtained with VASP for the all-aligned FA case, unless noted otherwise ($^{\dag}$ CP2K).}
\begin{tablenotes}
\footnotesize
\item \textbf{Legend:} 
$^{*}$ FA cations randomly oriented;  
DFT-D3: Grimme D3 correction with zero damping;  
DFT-D3(BJ): D3 with Becke-Johnson damping;  
rVV10: nonlocal van der Waals functional;  
$^{\dag}$ CP2K calculations;    
\end{tablenotes}
\label{tab:fasni3-S4-bandgaps}
\end{table}

With the 4$\times$4$\times$4* (random FA) structure established as the most physically realistic model, we next examine how different levels of theory influence its electronic structure. At the PBE level, calculations performed in Quantum ESPRESSO \cite{qe} produce a band gap that is strongly underestimated, consistent with the known behaviour of GGA functionals in Sn-based perovskites \cite{gga}. For this supercell, explicit SOC calculations were not feasible due to memory limitations, so we estimated the SOC contribution using a $\sim$0.30~eV correction obtained from converged calculations on smaller cells (see Supplementary Table~\ref{tab:fasni3_gaps}). Applying this correction yields a band gap of 0.49~eV, substantially lower than the experimental value of 1.41~eV \cite{exp_bg}, which is expected for PBE+SOC.

To access higher levels of theory, we turned to VASP \cite{vasp1, vasp2}, where hybrid functionals are more practical to apply. As in QE, we used zero-damping DFT-D3 for dispersion. The PBE band gaps obtained in VASP were consistent with those from QE (Supplementary Table~\ref{tab:code-suite-1a}), providing a reliable baseline for the hybrid calculations; the small differences arise from numerical settings and the pseudopotential choices in each code. Introducing exact exchange through PBE0 increases the band gap substantially, but unlike in FAPbI$_3$, where exact exchange and SOC tend to compensate \cite{PBE0}, no such cancellation occurs in FASnI$_3$. Applying PBE0+SOC to the 4×4×4* random-FA structure yields a gap of about 1.48 eV.

However, switching the dispersion scheme from DFT-D3 zero damping to the Becke-Johnson (BJ) \cite{BJ} variant lowers the gap by nearly 0.2~eV, an unexpectedly large difference for what should be a minor correction. The zero-damping form is known to overestimate short-range binding energies, and this tendency becomes problematic in systems such as $\alpha$-FASnI$_3$, where the stereochemically active Sn(II) lone pair introduces appreciable covalency into the Sn-I framework. Given this strong hybridization, empirical dispersion corrections can distort the balance between local bonding and long-range correlation, motivating the need for a nonlocal treatment of dispersion. 

To test this more systematically, we turned to CP2K \cite{CP2K}, which provides the non-empirical revised Vydrov-van Voorhis 2010 (rVV10) functional \cite{RVV} with self-consistent nonlocal correlation. Using the large 4$\times$4$\times$4* (random FA) supercell keeps finite-size and k-point artefacts small, so the remaining differences mainly reflect the functional and pseudopotential choices rather than the cell size. At the PBE0 level, switching from zero-damping D3 (as used in VASP) to rVV10 slightly relaxes the Sn-I framework and weakens the $\sigma$-overlap at the valence edge, raising the I-5$p$/Sn-5$s$ antibonding VBM by about 0.3~eV while leaving the CBM almost unchanged. This narrows the PBE0 gap from 2.07~eV to 1.76~eV. Adding the 0.30~eV SOC shift from converged QE calculations gives a final value of 1.46~eV, in close agreement with experiment. While VASP PBE0+SOC with zero-damping D3 may appear to match the experimental gap, PBE0+rVV10+SOC provides a more physically reliable description of the Sn-I bond lengths and the resulting band-edge positions.

\begin{table}[ht]
\centering
\resizebox{\textwidth}{!}{%
\renewcommand{\arraystretch}{1.2}
\setlength{\tabcolsep}{8pt}
\begin{tabular}{lcccc}
\hline
\textbf{Property} & \multicolumn{2}{c}{\textbf{4$\times$4$\times$4$^*$}} & \multicolumn{2}{c}{\textbf{6$\times$6$\times$6$^*$}} \\
\cline{2-5}
 & \makecell[c]{\textbf{Zero}\\\textbf{damping}} & \textbf{rVV10} & \makecell[c]{\textbf{Zero}\\\textbf{damping}} & \textbf{rVV10} \\
\hline
PJT Sn off-centering (\AA) & $0.32 \pm 0.09$  & $0.21 \pm 0.08 $ & $0.25 \pm 0.07$ & $0.15 \pm 0.06$ \\
Off-centering orientation order & 0.25 & 0.08 & 0.22 & 0.18  \\
Sn-I-Sn angle (°) & $163.84\pm 7.97$ & $162.73 \pm 8.40$ & $166.70 \pm 6.51$ & $166.80 \pm 6.23$ \\
Mean Squared Error (\AA) & 0.06 & 0.04 & 0.008 & 0.04  \\
Dipole Moment (Debye/ABX$_3$) & 0.36 & 0.73 & 0.60  & 0.44  \\
Octahedral anisotropy index & $0.13 \pm 0.03$ & $0.07 \pm 0.02$ & $0.09 \pm 0.02$ & $0.04 \pm 0.01$ \\
Normalized orbital overlap proxy & $0.85 \pm 0.08$ & $0.87 \pm 0.09$ & $0.91 \pm 0.07$  & $0.94 \pm 0.07$  \\
\hline
\end{tabular}%
}
\caption{Structural and electronic descriptors of $\alpha$-FASnI$_3$ as a function of supercell size and dispersion correction scheme. Reported values represent the mean ± standard deviation computed over all SnI$_6$ octahedra in each optimized structure.}
\label{tab:structure_metrics}
\end{table}

Looking at the structural and electronic metrics in Table~\ref{tab:structure_metrics}, rVV10 gives a noticeably more symmetric Sn-I environment than the D3 (zero-damping) version. Both the Sn off-centering amplitude and the octahedral anisotropy decrease, while the overall tilt pattern remains essentially unchanged. This indicates that the main effect of switching to rVV10 is on the local shape of the Sn potential rather than on the global tilt topology. In contrast, D3 (zero-damping) pulls certain Sn-I distances in too strongly, which deepens and skews the potential around Sn and drives larger static distortions. rVV10 distributes the correlation more evenly and consequently leaves the potential shallower and more isotropic, which naturally leads to smaller static displacements. The increase in the normalized Sn-I overlap index with rVV10 also points to slightly cleaner $s$-$p$ mixing at the valence edge, consistent with the $\sim$0.3~eV upward VBM shift reported earlier at the PBE0+rVV10 level. The behaviour of the macroscopic dipole is less straightforward: although rVV10 reduces both the magnitude and orientational order of Sn off-centerings, the total dipole moment increases slightly in the 4$\times$4$\times$4 supercell. This likely reflects FA dipoles reorganizing within a slightly less strained Sn-I lattice rather than a genuine increase in intrinsic polarization. The effect disappears in the 6$\times$6$\times$6 system, supporting the interpretation that it arises from finite-size or configurational effects rather than a real change in bulk behaviour.

Overall, rVV10 produces a more balanced Sn-I framework, while D3 (zero-damping) over-contracts specific Sn-I distances and drives larger static distortions. The rVV10 geometry therefore provides the more realistic structural model for $\alpha$-FASnI$_3$. However, finite-temperature calculations are still required to confirm that the structural and electronic trends identified here persist under thermal motion rather than being specific to 0 K.

\subsection*{Characterization of $\alpha$-FASnI$_3$ at 300~K}

\begin{figure}[htbp]
  \centering
  \includegraphics[width=1.0\textwidth]{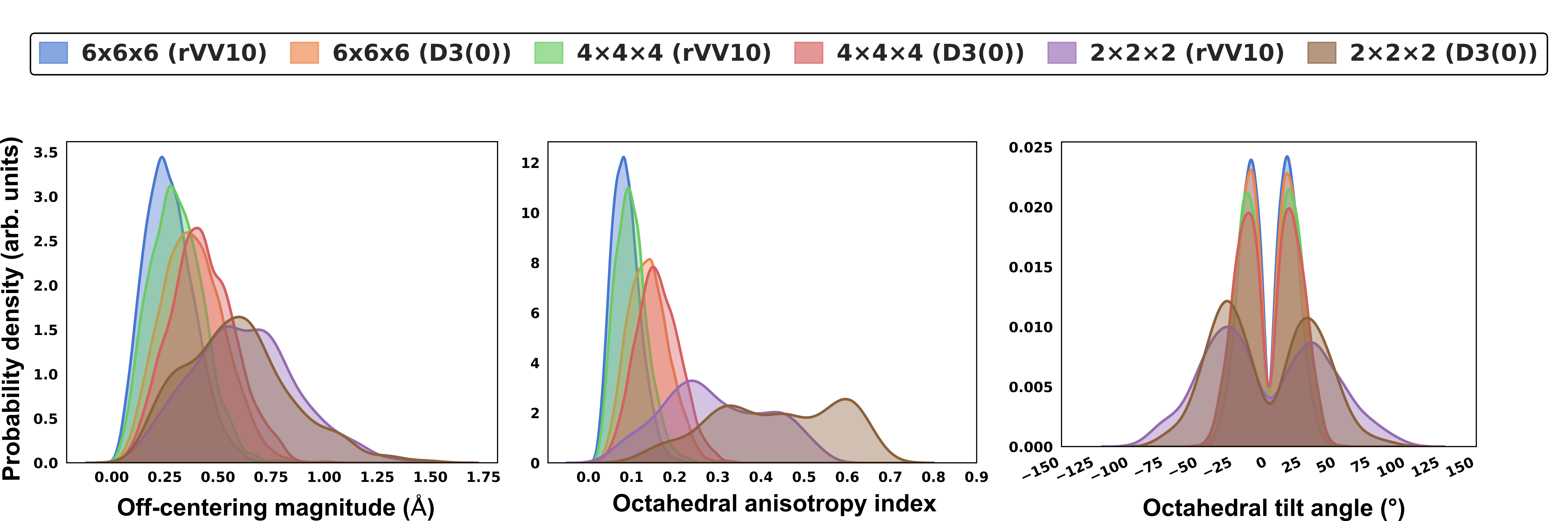}
  \caption{Distributions of Sn off-centering magnitudes (\si{\angstrom}), octahedral anisotropy indices, and signed tilt angles (°) from the last 2~ps of AIMD trajectories obtained with DFT-D3 and rVV10. The off-centering and anisotropy index are computed in the same manner as in the static 0~K analysis discussed earlier. The signed tilt ($-\theta$ / $+\theta$) denotes the deviation of Sn–I–Sn linkages from linearity.}
  \label{fig:comp_1}
\end{figure}

\begin{table}[h!]
\centering
\renewcommand{\arraystretch}{1.3}
\setlength{\tabcolsep}{10pt}
\begin{tabular}{
    >{\centering\arraybackslash}m{3cm}
    >{\centering\arraybackslash}m{2cm}
    >{\centering\arraybackslash}m{2cm}
    >{\centering\arraybackslash}m{2cm}
    >{\centering\arraybackslash}m{2cm}
}
\toprule
\textbf{System} & \multicolumn{2}{c}{\textbf{PBE}} & \multicolumn{2}{c}{\textbf{PBE0}} \\ 
\cmidrule(lr){2-3} \cmidrule(lr){4-5}
 & \textbf{D3} & \textbf{rVV10} & \textbf{D3} & \textbf{rVV10} \\ 
\midrule
1$\times$1$\times$1 &3.70$\pm$0.22 & 3.72$\pm$0.20 & 6.18$\pm$0.32  & 5.51$\pm$0.21 \\
2$\times$2$\times$2 & 2.05$\pm$0.14  & 1.82$\pm$0.11  &  2.56$\pm$0.56 & 2.67$\pm$0.18  \\
4$\times$4$\times$4 & 1.19$\pm$0.05  & 0.98$\pm$0.05  & 2.16$\pm$0.07 &  1.90$\pm$0.05\\
6$\times$6$\times$6 & 1.05$\pm$0.04 & 0.89$\pm$0.03 &1.94$\pm$0.03 & 1.76$\pm$0.03 \\
\bottomrule
\end{tabular}
\caption{Comparison of PBE and PBE0 calculations with D3 and rVV corrections across different system sizes.}
\label{tab:pbe-pbe0-comparison}
\end{table}
\begin{figure}[t]
  \centering
  \includegraphics[width=1.0\textwidth]{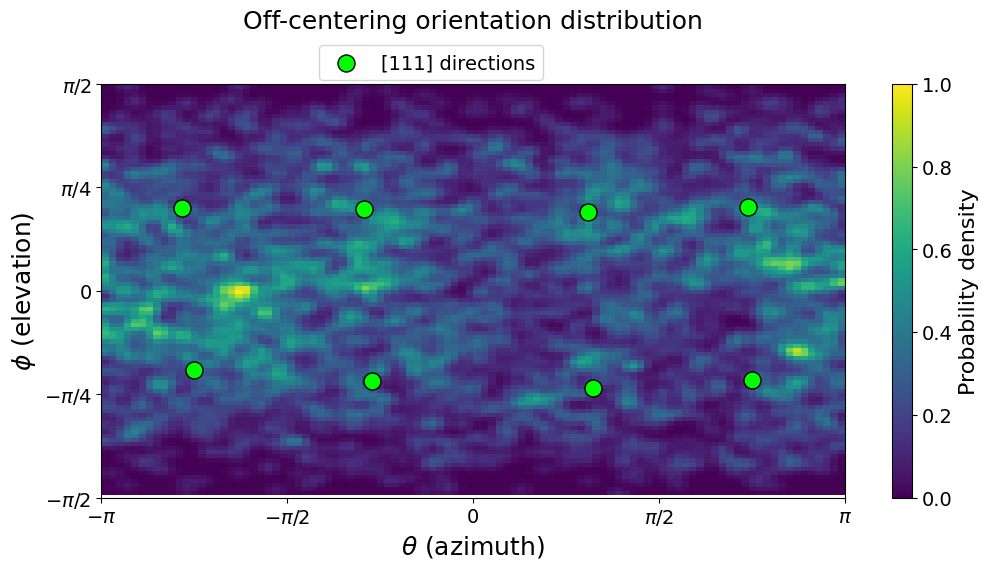}
\caption{Distribution of instantaneous Sn off-centering vectors, computed from the last 2~ps of the 300~K trajectory for the 4$\times$4$\times$4$^*$ random-FA supercell. Probability density is plotted over polar ($\phi$) and azimuthal ($\theta$) angles. The eight crystallographic $\langle 111\rangle$ directions are marked in green.}

  \label{fig:offcentering_heatmap}
\end{figure}

Ab initio molecular dynamics (AIMD) simulations at 300~K were carried out for $\alpha$-FASnI$_3$ in the NPT ensemble using the same setup as for $\alpha$-FAPbI$_3$. Both PBE and PBE0 levels were explored with DFT-D3 or rVV10 dispersion corrections. The FA cations exhibited dynamic orientational disorder at finite temperature, and thermal averages of key structural and electronic descriptors were extracted from equilibrated trajectory snapshots. For larger supercells, PBE0 single-point evaluations on PBE-AIMD snapshots were used to include thermal effects efficiently. Further computational and simulation details are described in the Discussion section.

At 300 K, the same trends seen at 0 K remain. In the largest system we simulated (the 6×6×6 supercell), PBE0 with DFT-D3 (zero damping) and the converged SOC correction gives a band gap of about 1.6 eV, which overshoots experiment. Switching to the non-local rVV10 treatment lowers the gap to ~1.46 eV, bringing it into the experimental range. Notably, we had already reached this level of agreement at 0 K using the 4×4×4* cell; however, at finite temperature the additional structural fluctuations appear to require a larger supercell for the band edge positions to converge.
  
Extending the analysis to 300~K basically continues the 0~K picture. As shown in Fig.~\ref{fig:comp_1} and quantified in Supplementary Table \ref{struc_param_300K}, DFT-D3 still gives slightly larger mean Sn off-centering and octahedral anisotropy than rVV10, but the thermal motion is now large enough that their distributions overlap almost completely. So the functional-dependent differences are still there, just much less pronounced once vibrations are included. Even so, rVV10 behaves in the same way as before: the local Sn-I potential stays a bit smoother and more symmetric, and the lattice remains marginally softer and more isotropic compared to D3. The octahedral tilt distributions, with their characteristic bimodal in-phase/out-of-phase peaks, remain essentially unchanged between the two functionals. This again shows that rVV10 mainly affects the local Sn-I distortions without altering the overall tilt topology, consistent with what we found at 0~K.

To analyse the Sn displacements at finite temperature, we projected the instantaneous off-centering vectors onto spherical angles (Fig.~\ref{fig:offcentering_heatmap}). The distribution is concentrated along the eight $\langle111\rangle$ directions, a behaviour also observed in other Sn(II) halide perovskites \cite{PJTE_2}. In $\alpha$-FASnI$_3$, this pattern persists at 300~K, showing that the preferred off-centering directions survive thermal motion. This follows from the second-order Jahn-Teller mechanism, where mixing between the Sn 5$s^2$ lone pair and antibonding Sn-5$p$/I-5$p$ states creates eight equivalent minima along $\langle111\rangle$.

To assess the finite-temperature behaviour of the Sn lone pair, we extracted its rotational frequency from the 300 K trajectory, obtaining a value of 72.8 cm$^{-1}$ (see Discussion for methodological details). This falls within the mid-frequency range reported in Raman spectra of CsSnBr$_3$ , where such modes are associated with local distortions driven by lone-pair activity \cite{wannier}. This indicates that the Sn 5$s^2$ lone pair in $\alpha$-FASnI$_3$ remains stereochemically active at room temperature and contributes dynamically to local structural fluctuations.

\section*{Discussion}

In conclusion, our results show that obtaining a physically reliable description of $\alpha$-FASnI$_3$ requires modelling choices that explicitly reflect the behaviour of Sn(II), whose stereochemically active lone pair drives local PJT distortions. Small or symmetry-restricted cells suppress this physics and introduce artefacts from FA dipoles and periodic boundary conditions. By systematically increasing the system size, we find that a $4 \times 4 \times 4^{\ast}$ supercell with random FA orientations is the smallest model that removes macroscopic fields, avoids periodic coherence in the Sn-I network, restores cubic average symmetry, and yields a size-independent Sn off-centering amplitude. This structure therefore serves as a physically meaningful 0~K reference for $\alpha$-FASnI$_3$.

On the electronic side, PBE+SOC severely underestimates the band gap, and hybrid functionals require an appropriate dispersion treatment to avoid overstabilising short Sn-I contacts. Among the methods tested, PBE0 combined with the nonlocal rVV10 scheme gives the most balanced description of the Sn-I interaction. For the $4 \times 4 \times 4^{\ast}$ model, PBE0+rVV10 together with the SOC correction extracted from converged QE calculations yields a band gap of 1.46~eV, in close agreement with experiment. Finite-temperature AIMD further shows that the Sn off-centering remains local and retains its $\langle 111 \rangle$ preference once FA dipoles cancel, and that recovering the experimental gap at 300~K requires both hybrid exchange and nonlocal dispersion in a sufficiently large ($6 \times 6 \times 6$) supercell. Overall, these results provide a consistent modelling framework for $\alpha$-FASnI$_3$ and outline the key ingredients needed to capture its structure, lone-pair-driven distortions, and band-edge physics.

\subsection*{First-Principles DFT Simulations at 0 K}

DFT simulations at 0 K were carried out using Quantum Espresso (QE) \cite{qe}, VASP \cite{vasp1, vasp2}, and CP2K \cite{CP2K} to take advantage of the strengths of each code suite. QE and VASP, both plane-wave-based frameworks, were used for systematic comparisons across different levels of theory, while CP2K, based on the Gaussian plane-wave (GPW) approach, offered computational efficiency for larger systems. In Quantum Espresso, fully relativistic norm-conserving PBE pseudopotentials from the DOJO library \cite{dojo1, dojo2} were used with energy cutoffs of 150 Ry for wavefunctions and 600 Ry for charge density. In VASP, projector-augmented wave (PAW) \cite{vasp2} PBE pseudopotentials were employed with a cutoff of 600 eV. Both codes used Γ-centered Monkhorst-Pack k-point grids \cite{kgrid} and included van der Waals interactions through the DFT-D3 (zero damping) correction \cite{D3}; additional tests in VASP were performed using the DFT-D3(BJ) scheme \cite{BJ}. Higher-level single-point calculations beyond PBE and SOC corrections were carried out on the fully relaxed PBE structures. In CP2K, DFT calculations were performed within the GPW framework using Goedecker-Teter-Hutter (GTH) pseudopotentials \cite{GTH} and a polarized double-zeta Gaussian basis set (DZVP-MOLOPT) \cite{DZVP} for valence electrons. The energy cutoff for the plane-wave expansion of the electron density was set to 600 Ry at the Gamma point. Dispersion interactions were treated using both DFT-D3 \cite{D3} (zero damping) and rVV10 schemes \cite{RVV} at the PBE and PBE0 levels of theory, while SOC corrections were adopted from the well-converged Quantum Espresso runs, as these effects are local and converge consistently across systems.

When comparing results across Quantum Espresso, VASP, and CP2K, differences in absolute band gap values are expected because of variations in pseudopotentials, basis sets, and numerical implementations (see Supplementary Tables \ref{tab:code-suite-1a}, \ref{tab:vasp-qe-comparison}). What stands out more clearly, however, is the effect of structural differences between the codes. When the self-consistent runs in VASP are performed using the PBE-optimized structures from Quantum Espresso and compared with those using structures optimized directly in VASP, the largest difference appears for the \(2 \times 2 \times 2\) supercell. This suggests that the two relaxations converge to different local minima. On closer inspection, VASP tends to enforce a more cubic symmetry during variable-cell relaxation, which is not appropriate for such a small cell where distortions are expected to compensate the dipoles from the aligned FA cations. For the larger \(4 \times 4 \times 4\) cell, where the structure naturally relaxes without artificial symmetry constraints, the discrepancy becomes much smaller. This shows that even to make meaningful comparisons across frameworks, it is essential to start from a physically correct relaxed structure.

\subsection*{Ab Initio Molecular Dynamics at 300 K}

Ab initio molecular dynamics simulations at 300~K were performed within the DFT framework implemented in CP2K, following the same computational setup as for the static 0~K calculations. Simulations were carried out using Born-Oppenheimer dynamics in the NPT flexible cell ensemble with a time step of 0.5~fs. Each trajectory was propagated for 6-10~ps at both the PBE and PBE0 levels of theory, employing either the DFT-D3 (zero damping) \cite{D3} or the non-local rVV10 dispersion correction \cite{RVV}. The temperature was controlled by the Bussi thermostat \cite{Bussi} and the pressure by the Martyna barostat \cite{Martyna}. For the larger 4$\times$4$\times$4 and 6$\times$6$\times$6 supercells, full AIMD simulations at the PBE0 level were computationally demanding; therefore, representative snapshots from the PBE trajectories were used for single-point self-consistent field (SCF) calculations with PBE0. This approach captures thermal structural effects from PBE while incorporating the improved electronic description of PBE0.A systematic investigation was carried out from the 1$\times$1$\times$1 to the
6$\times$6$\times$6 supercell. Band gaps were extracted from the instantaneous projected density of states (PDOS) over the equilibrated final $\sim$2~ps and reported as time-averaged values.
Structural and dipole moment properties were similarly averaged over the same period.

\subsection*{Description of Structural Properties}

The structural descriptors were obtained from the SnI$_6$ octahedra and Sn-I-Sn bridges identified in each configuration. For each simulation cell, these quantities were evaluated by averaging over all Sn-centered octahedra, with the standard deviation reflecting the variability within the structure. The PJT Sn off-centering corresponds to the displacement of the Sn atom from the geometric center of its surrounding I$_6$ cage, expressed as a vector magnitude in~\AA. The off-centering orientation order quantifies the degree of collective alignment of these displacements, with a value of~1 corresponding to perfect alignment and~0 to random orientations. The Sn-I-Sn angle represents the tilt between neighboring octahedra, while the octahedral anisotropy index is defined as the standard deviation of the six Sn-I bond lengths normalized by their mean, describing deviations from an ideal octahedral geometry. From these geometries, an effective orbital overlap proxy was calculated for each Sn-I-Sn bridge based on the bond angle~($\theta$) and Sn-I distances~($r_1$, $r_2$), and reported in normalized form to allow comparison across cells and conditions. At~0~K, the reported values correspond to the final relaxed structure, as listed in Tables~\ref{fasni3_structural} and~\ref{tab:structure_metrics}. The Supporting Information (Supplementary Note 4) mentions the formulas used to compute these quantities.

\subsection*{Wannier-Center Analysis of Lone-Pair Rotational Dynamics}

To characterize the rotational dynamics of the stereochemically active Sn\,5s$^2$ lone
pair in $\alpha$-FASnI$_3$, we performed a Wannier-center analysis on a
4$\times$4$\times$4 supercell equilibrated at 300~K. A representative NPT-F snapshot
was propagated for $\sim$20~ps in the NVE ensemble, during which maximally localized
Wannier centers (MLWCs) were saved every 50~fs. For each frame, the Wannier center
lying within 0.3\,\AA\ of a given Sn atom was identified as the corresponding
lone-pair Wannier center (LP-WC).

The instantaneous lone-pair orientation was defined by the normalized vector from each
Sn atom to its associated LP-WC. These unit vectors were tracked over the full
trajectory and used to compute the second-order Legendre rotational time-correlation
function. The decay of this function yields the rotational correlation time, from
which a rotational frequency is obtained. Ensemble averages were taken over all Sn sites and all valid time origins. The Supporting Information (Supplementary Note 5) mentions the formulas used to compute these quantities.

\newpage

\printbibliography

\section*{Acknowledgements}
U.R. acknowledges support from the Swiss National Science Foundation (Grant No.~200020\_219440) and computational resources provided by the Swiss National Supercomputing Centre (CSCS). V.C. acknowledges computational resources provided by the Swiss National Supercomputing Centre (CSCS).

\section*{Author Information}

\subsection*{Authors and Affiliations}

\textbf{Laboratory of Computational Chemistry and Biochemistry, 
Institute of Chemical Sciences and Engineering, 
Swiss Federal Institute of Technology (EPFL), Lausanne, Switzerland} \\
Mridhula Venkatanarayanan, Vladislav Slama, 
Madhubanti Mukherjee, Andrea Vezzosi,
Ursula Rothlisberger \& Virginia Carnevali

\subsection*{Contributions}
M.V. and V.A. conceived the idea. M.V. and V.A. performed the theoretical simulations. 
M.V., V.S., M.M., and A.V. wrote the manuscript under the supervision of V.C. and U.R. 
All authors contributed to the discussion and to the writing of the manuscript.

\subsection*{Corresponding authors}

Ursula Rothlisberger (\href{mailto:ursula.roethlisberger@epfl.ch}{ursula.roethlisberger@epfl.ch}) \\[2pt]
Virginia Carnevali (\href{mailto:virginia.carnevali@epfl.ch}{virginia.carnevali@epfl.ch})

\section*{Ethics declarations}

\subsection*{Competing interests}
The authors declare no competing interests.

\end{document}

\sloppy

\begin{titlepage}
\pagenumbering{gobble}
\centering
\onehalfspacing        

{\Large Supplementary Information\par}
\vspace{1.5em}

{\LARGE Coupled Structural and Electronic Requirements in $\alpha$-FASnI$_3$ Imposed by the Sn(II) Lone Pair
\par}

\vspace{2em}

{\large
Mridhula Venkatanarayanan,
Vladislav Slama,
Madhubanti Mukherjee,
Andrea Vezzosi,
Ursula Rothlisberger$^{*}$, and Virginia Carnevali$^{\dagger}$\par}

\medskip
\textit{Laboratory of Computational Chemistry and Biochemistry, 
Institute of Chemical Sciences and Engineering, 
Swiss Federal Institute of Technology (EPFL), Lausanne, Switzerland}

\medskip
$^{*}${ursula.roethlisberger@epfl.ch} \\
$^{\dagger}${virginia.carnevali@epfl.ch}

\vspace{2em}

\vfill 
\end{titlepage}

\pagenumbering{arabic} 
\setstretch{1.25}

\section*{Supplementary Note 1a. Size and method dependence of the FASnI$_3$ band gap}

\begin{table}[h!]
\centering
\small
\setlength{\tabcolsep}{4pt}
\renewcommand{\arraystretch}{1.2}
\begin{threeparttable}
\caption[Band-gap convergence in FASnI$_3$]{Kohn-Sham band-gap convergence in FASnI$_3$ as a function of supercell size, 
\textit{k}-point sampling, temperature (0~K relaxations and 300~K NPT-F$^{\ddagger}$ averages), 
and level of theory (PBE and PBE0, with and without SOC). 
At 300~K, each trajectory is propagated for 6-10~ps, and the band gaps are obtained by 
averaging the instantaneous projected density of states over the final $\sim$2~ps of the run.}

\label{tab:fasni3_gaps}
\begin{tabular}{@{}lcccccccc@{}}
\toprule
\multirow{2}{*}{\makecell[l]{Simulation\\cell}} &
\makecell{NPT-F$^{\dag}$\\PBE\\300 K} &
\makecell{NPT-F$^{\dag}$\\PBE0\\300 K} &
\makecell{relax\\PBE\\0 K} &
\makecell{vc-relax\\PBE\\0 K} &
\makecell{vc-relax\\PBE+SOC\\0 K} &
\makecell{vc-relax\\PBE0\\0 K} &
\makecell{vc-relax\\PBE0+SOC\\0 K} &
\makecell{$n\times n\times n$\\k-point grid} \\
\midrule

\multirow[t]{6}{*}{12-atom}
& \(3.69 \pm 0.18\) & \(6.18 \pm 0.32\) & 3.68 & 3.74 & 3.50 & 5.83 & 5.57 & 1 \\
&  &  & 1.69 & 2.69 & 2.52 & 4.29 & 4.10 & 2 \\
&  &  & 0.76 & 0.91 & 0.69 & 2.73 & 2.49 & 4 \\
&  &  & 0.72 & 0.89 & 0.60 & 2.70 & 2.39 & 6 \\
&  &  & 0.71 & 0.86 & 0.57 & 2.67 & 2.36 & 8 \\
&  &  & 0.71 & 0.86 & 0.57 & 2.65 & (2.36) & 10 \\

\midrule[\heavyrulewidth] 

\multirow[t]{5}{*}{96-atom}
& \(2.04 \pm 0.14\) & \(2.56 \pm 0.56\) & 1.57 & 2.15 & 2.05 & 3.36 & 3.25 & 1 \\
&  &  & 0.76 & 0.87 & 0.63 & 2.04 & 1.76 & 2 \\
&  &  & 0.72 & 0.86 & 0.56 & 2.02 & (1.72) & 4 \\
&  &  & 0.71 & 0.84 & 0.55 & 1.99 & (1.69) & 6 \\
&  &  & 0.80 & 0.70 & 0.39 & - & - & 6$^{*}$ \\

\midrule[\heavyrulewidth]

\multirow[t]{3}{*}{768-atom}
& \(1.35 \pm 0.07\) & \(2.16 \pm 0.07\) & 0.76 & 0.89 & (0.59) & - & - & 1 \\
&  &  & 0.93 & 0.92 & (0.62) & - & - & 1$^{*}$ \\
&  &  & 0.72 & 0.79 & (0.49) & - & - & 2 \\

\midrule[\heavyrulewidth]

\multirow[t]{1}{*}{2592-atom}
& \(1.18 \pm 0.01\) & \(1.94 \pm 0.03\) & - & - & - & - & - & 1 \\

\bottomrule
\end{tabular}

\footnotesize

\textit{Legend:}\\
$^{*}$\; FA randomly oriented.\\
$^{\ddagger}$\; CP2K.\\
(no symbol)\; QE + DFT-D3 (Grimme).\\
( )\; SOC correction from converged (PBE+SOC)-PBE.\\
A dash (-) indicates not computed due to memory bottleneck.

\end{threeparttable}
\end{table}

\newpage

\section*{Supplementary Note 1b. Size and method dependence of the FASnI$_3$ band edges}
\begin{table}[h!]
\centering
\caption{Kohn-Sham band edges (VBM: valence band maximum; CBM: conduction band minimum) reported for different simulation cells, levels of theory, and structural relaxation schemes.}
\renewcommand{\arraystretch}{1.15}
\setlength{\tabcolsep}{3.5pt}
\small 
\label{tab:fasni3_edges}
\begin{adjustbox}{max width=\textwidth}
\begin{tabularx}{\textwidth}{
    >{\raggedright\arraybackslash}p{3.0cm}
    |*{2}{Y}|*{2}{Y}|*{2}{Y}|*{2}{Y}|*{2}{Y}|c}
\Xhline{1.1pt}
\makecell{Simulation\\[-1pt]cell} &
\multicolumn{2}{c|}{\makecell{relax\\ PBE\\ 0 K\\ (eV)}} &
\multicolumn{2}{c|}{\makecell{vc-relax\\ PBE\\ 0 K\\ (eV)}} &
\multicolumn{2}{c|}{\makecell{vc-relax\\ PBE+SOC\\ 0 K\\ (eV)}} &
\multicolumn{2}{c|}{\makecell{vc-relax\\ PBE0\\ 0 K\\ (eV)}} &
\multicolumn{2}{c|}{\makecell{vc-relax\\ PBE0+SOC\\ 0 K\\ (eV)}} &
\makecell{$n\times n\times n$\\ k-point\\ grid} \\
\hline
\multicolumn{1}{c|}{} &
VBM & CBM &
VBM & CBM &
VBM & CBM &
VBM & CBM &
VBM & CBM &
\\
\Xhline{0.8pt}

\multirow{6}{*}{\makecell{12-atoms\\ (1 $\times$ 1 $\times$ 1)}} 
& 1.52 & 5.19 & 2.45 & 6.19 & 2.69 & 6.19 & 1.37 & 7.20 & 1.63 & 7.20 & 1 \\
& 2.54 & 4.23 & 1.43 & 4.12 & 1.57 & 4.09 & 0.46 & 4.75  & 0.60 & 4.70 & 2 \\
& 2.96 & 3.72 & 2.87 & 3.78 & 2.97 & 3.66 & 1.80 & 4.53 & 1.90 & 4.39 & 4 \\
& 2.98 & 3.69 & 2.97 & 3.86 & 3.08 & 3.69 & 1.93 & 4.63 & 2.05 & 4.44 & 6 \\
& 2.98 & 3.69 & 3.02 & 3.88 & 3.13 & 3.70 & 1.98 & 4.64 & 2.10 & 4.45 & 8 \\
& 2.98 & 3.69 & 3.03 & 3.89 & 3.14 & 3.71 & 1.99 & 4.64 & - & - & 10 \\
\Xhline{1.1pt}
\multirow{5}{*}{\makecell{96-atoms\\ (2 $\times$ 2 $\times$ 2)}} 
& 2.50 & 4.06 & 2.17 & 4.32 & 2.27 & 4.32 & 1.44 & 4.80 & 1.54 & 4.79 & 1 \\
& 2.97 & 3.72 & 2.90 & 3.77 & 3.00 & 3.63 & 2.16 & 4.20 & 2.27 & 4.03 & 2 \\
& 2.97 & 3.69 & 3.03 & 3.90 & 3.14 & 3.70 & 2.31 & 4.33 & -  & - & 4 \\
& 2.98 & 3.69 & 3.03 & 3.87 & 3.15 & 3.70 & 2.31 & 4.30 &- & - & 6 \\
& 2.94$^{*}$ & 3.75$^{*}$ & 3.21$^{*}$ & 3.91$^{*}$& 3.32$^{*}$ & 3.71$^{*}$ & - & - & - & - & 6$^{*}$ \\
\Xhline{1.1pt}

\multirow{3}{*}{\makecell{768-atoms\\ (4 $\times$ 4 $\times$ 4)}} 
& 2.97 & 3.72 & 2.88 & 3.77 & - & - & - & - & - & - & 1 \\
& 2.91$^{*}$ & 3.84$^{*}$ & 2.96$^{*}$  & 3.87$^{*}$  & - & - & - & - & - & - & 1$^{*}$ \\
& 2.97 & 3.69 & 3.07  & 3.86  & - & - & - & - & - & - & 2 \\
\bottomrule
\end{tabularx}
\end{adjustbox}

\vspace{2pt}
\raggedright\footnotesize
\textit{Legend:}\\
$^{*}$\; FA randomly oriented.\\
$^{\ddagger}$\; CP2K.\\
(no symbol)\; QE + DFT-D3 (Grimme).\\
( )\; SOC correction from converged (PBE+SOC)-PBE.\\
A dash (-) indicates not computed due to memory bottleneck.
\end{table}

\newpage

\section*{Supplementary Note 2a. Structural Parameters and Dipole Moment}
\begin{table}[h!]
\centering
\renewcommand{\arraystretch}{1.3}
\setlength{\tabcolsep}{8pt}
\caption{Mean Squared Error (MSE) of lattice vectors and dipole moments (in Debye per ABX$_3$) 
for different simulation cell sizes. At 0\,K, values are reported for Random and All-aligned (AA) configurations. 
At 300\,K, results correspond to average over the last 2 ps (PBE).}
\label{mse_dip}
\begin{tabular}{lcccccc}
\toprule
\multirow{3}{*}{\textbf{Simulation Cell}}
  & \multicolumn{3}{c}{\textbf{MSE}}
  & \multicolumn{3}{c}{\textbf{Dipole Moment (Debye/ABX$_3$)}} \\
\cmidrule(lr){2-4}\cmidrule(lr){5-7}
  & \multicolumn{2}{c}{\textbf{0 K}} & \textbf{300 K}
  & \multicolumn{2}{c}{\textbf{0 K}} & \textbf{300 K (PBE)} \\
\cmidrule(lr){2-3}\cmidrule(lr){5-6}
  & Random & AA & 
  & Random & AA & \\
\midrule
12 atom   & \multicolumn{2}{c}{0.02} & 0.05 $\pm$ 0.03
          & \multicolumn{2}{c}{0.36} & 2.46 $\pm$ 1.02 \\
96 atom   & 1.65 & 2.22 & 2.44 $\pm$ 0.25
          & 2.11 & 3.32 & 4.08 $\pm$ 0.56 \\
768 atom  & 0.06 & 2.01 & 0.16 $\pm$ 0.03
           & 0.36 & 0.94 & 1.00 $\pm$ 0.33 \\
2592 atom & 0.008 & 3.79 & 0.22 $\pm$ 0.04
          & 0.60 & 0.49 & 0.36 $\pm$ 0.01 \\
\bottomrule
\end{tabular}
\end{table}

\section*{Supplementary Note 2b. Structural parameters at 300 K (PBE)}

\begin{table}[h!]
\centering
\caption{Mean $\pm$ standard deviation of the structural parameters over all SnI$_6$ cages, averaged over the last 2~ps of the 300~K PBE AIMD trajectories.}
\label{struc_param_300K}
% Increase horizontal spacing between columns
\setlength{\tabcolsep}{12pt}
\begin{tabular}{lcccc}
\toprule
\textbf{Simulation cell} 
& \multicolumn{2}{c}{\textbf{Sn off-centering (Å)}} 
& \multicolumn{2}{c}{\textbf{Octahedral anisotropy}} \\
\cmidrule(lr){2-3} \cmidrule(lr){4-5}
& D3(0) & rVV10 & D3(0) & rVV10 \\
\midrule
2$\times$2$\times$2 & \(0.59 \pm 0.25\) & \(0.62 \pm 0.24\) & \(0.44 \pm 0.15\) & \(0.30 \pm 0.12\) \\
4$\times$4$\times$4 & \(0.43 \pm 0.15\) & \(0.31 \pm 0.13\) & \(0.16 \pm 0.05\) &  \(0.10 \pm 0.04\)\\
6$\times$6$\times$6 & \(0.38 \pm 0.15\)& \(0.27 \pm 0.12\) & \(0.14 \pm 0.05\) & \(0.09 \pm 0.03\) \\
\bottomrule
\end{tabular}
\end{table}

\newpage
\section*{Supplementary Note 3a. Cross-code comparison of band gaps}

\begin{table*}[htbp]
\centering
\caption{Band gaps at 0~K computed with the PBE functional across several supercell sizes, comparing plane-wave (PW) and Gaussian–plane-wave (GPW) frameworks. PW results include Quantum~Espresso (QE) with norm-conserving pseudopotentials (NCPP) and QE with PAW, the latter included to enable direct comparison with VASP (PAW-only). GPW calculations were performed with CP2K. Two optimisation protocols are shown: relax (geometry optimisation at fixed lattice parameters) and vc-relax (simultaneous optimisation of cell and geometry). Structures marked with * correspond to random FA orientations.
}
\label{tab:code-suite-1a}
\small
\begin{tabular}{
l
S[table-format=1.2] S[table-format=1.2] 
S[table-format=1.2] S[table-format=1.2]  
S[table-format=1.2] S[table-format=1.2]  
@{\hspace{1.2em}}                        
S[table-format=1.2] S[table-format=1.2]  
@{\hspace{1.2em}}                        
c                                         
}
\toprule
& \multicolumn{6}{c}{\textbf{PW}} &
  \multicolumn{2}{c}{\textbf{GPW}} &
  \textbf{k-grid} \\
\cmidrule(lr){2-7}\cmidrule(lr){8-9}\cmidrule(lr){10-10}
& \multicolumn{2}{c}{\makecell{QE \\ (NCPP+NLCC)}} &
  \multicolumn{2}{c}{\makecell{QE \\ (PAW)}} &
  \multicolumn{2}{c}{\makecell{VASP \\ (PAW)}} &
  \multicolumn{2}{c}{\makecell{CP2K}} &
  \\
\cmidrule(lr){2-3}\cmidrule(lr){4-5}\cmidrule(lr){6-7}\cmidrule(lr){8-9}
\textbf{Simulation cell} &
\multicolumn{1}{c}{relax} & \multicolumn{1}{c}{vc-relax} &
\multicolumn{1}{c}{relax} & \multicolumn{1}{c}{vc-relax} &
\multicolumn{1}{c}{relax} & \multicolumn{1}{c}{vc-relax} &
\multicolumn{1}{c}{relax} & \multicolumn{1}{c}{vc-relax} &
\\
\midrule

\multirow{6}{*}{\makecell{$1\times1\times1$ \\ (12 atoms)}} &
3.68 & 3.74 & 3.70 & 3.76 & 3.67 & 3.74 &
\multicolumn{1}{c}{ } & \multicolumn{1}{c}{ } &
$1\times1\times1$ \\

& 1.69 & 2.69 & 1.63 & 2.67 & 3.76 & 3.72 &
\multicolumn{1}{c}{-} & \multicolumn{1}{c}{-} &
$2\times2\times2$ \\

& 0.76 & 0.91 & 0.74 & 0.94 & 2.02 & 2.02 &
\multicolumn{1}{c}{-} & \multicolumn{1}{c}{-} &
$4\times4\times4$ \\

& 0.72 & 0.89 & 0.70 & 0.92 & 1.39 & 1.36 &
\multicolumn{1}{c}{-} & \multicolumn{1}{c}{-} &
$6\times6\times6$ \\

& 0.71 & 0.86 & 0.70 & 0.85 & 1.11 & 1.11 &
\multicolumn{1}{c}{-} & \multicolumn{1}{c}{-} &
$8\times8\times8$ \\

& 0.71 & 0.86 & 0.69 & 0.88 & 0.97 & 0.97 &
\multicolumn{1}{c}{-} & \multicolumn{1}{c}{-} &
$10\times10\times10$ \\
\midrule

\multirow{4}{*}{\makecell{$2\times2\times2$ \\ (96 atoms)}} &
1.57 & 2.15 & 1.69 & 2.66 & 1.41 & 1.99 &  
1.73 & 2.42 &
$1\times1\times1$ \\

& 0.76 & 0.87 & 0.75 & 1.09 & 2.04 & 2.02 &
\multicolumn{1}{c}{-} & \multicolumn{1}{c}{-} &
$2\times2\times2$ \\

& 0.72 & 0.86 & 0.70 & 0.86 & 1.11 & 1.11 &
\multicolumn{1}{c}{-} & \multicolumn{1}{c}{-} &
$4\times4\times4$ \\

& 0.71 & 0.84 & 0.69 & 0.84 & \multicolumn{1}{c}{-} & \multicolumn{1}{c}{-} &
\multicolumn{1}{c}{-} & \multicolumn{1}{c}{-} &
$6\times6\times6$ \\
\midrule

\multirow{2}{*}{\makecell{$4\times4\times4$ \\ (768 atoms)}} &
0.76 & 0.89 & \text{-} & \text{-} & \text{-} & 0.98 &  
0.84 & 1.24 &
$1\times1\times1$ \\

&\multicolumn{1}{c}{0.93} & \multicolumn{1}{c}{0.92} &
\multicolumn{1}{c}{-} & 
\multicolumn{1}{c}{-} &
\multicolumn{1}{c}{-} & 
\multicolumn{1}{c}{0.81} &
\multicolumn{1}{c}{1.00} & \multicolumn{1}{c}{1.09} &
$1\times1\times1^{*}$ \\

\midrule
\bottomrule
\end{tabular}
\end{table*}

\newpage 
\section*{Supplementary Note 3b. Cross-code comparison of band gaps}

\begin{table}[h!]
\centering

\caption{Band gaps (in eV) at 0~K calculated in VASP for multiple simulation cells and levels of theory. 
Two types of calculations are shown: ($\phi$) SCF calculations starting from the PBE-optimized (vc-relax) structure in VASP, used consistently across all levels of theory; 
($\psi$) SCF calculations starting from the PBE-optimized structure obtained in QE. Structures marked with * correspond to random FA orientations. }
\label{tab:vasp-qe-comparison}
\setlength{\tabcolsep}{6pt}
\renewcommand{\arraystretch}{1.2}
\begin{adjustbox}{width=\textwidth}
\begin{tabular}{llccccccc}
\toprule
\multirow{2}{*}{\textbf{Simulation cell}}
& \multicolumn{6}{c}{\textbf{Functional}} 
& \multirow{2}{*}{\begin{tabular}[c]{@{}c@{}}\textbf{k-grid}\\ \textbf{($n\times n\times n$)}\end{tabular}}
 \\
\cmidrule(lr){2-7}
& \textbf{PBE} & \textbf{PBE+SOC} & \textbf{PBE0} & \textbf{PBE0+SOC} & \textbf{HSE06} & \textbf{HSE06+SOC} & \\
\midrule
\multirow{2}{*}{\begin{tabular}[c]{@{}c@{}}$1\times1\times1$ \\ (12 atoms)\end{tabular}} 
  & 0.97$^{\phi}$ & 0.70$^{\phi}$ & 1.87$^{\phi}$ & 1.52$^{\phi}$ & 1.35$^{\phi}$ & 1.04$^{\phi}$  & \multirow{2}{*}{$10\times10\times10$} \\
  & 1.06$^{\psi}$  & 0.77$^{\psi}$  & 2.02$^{\psi}$  & 1.66$^{\psi}$  & 1.48$^{\psi}$  & 1.15$^{\psi}$  &  \\
\addlinespace
\multirow{2}{*}{\begin{tabular}[c]{@{}c@{}}$2\times2\times2$ \\ (96 atoms)\end{tabular}} 
  & 0.89$^{\phi}$ & 0.61$^{\phi}$ & - & - & - & - & \multirow{3}{*}{$6\times6\times6$} \\
  & 0.11$^{\psi}$  & 0.06$^{\psi}$  & - & - & - & - & \\
\addlinespace

\multirow{2}{*}{\begin{tabular}[c]{@{}c@{}}$4\times4\times4$* \\ (768 atoms)\end{tabular}} 
  & 0.81$^{\phi}$ & 0.49$^{\phi}$ & 1.84$^{\phi}$ & 1.48$^{\phi}$ & 1.25$^{\phi}$ &  0.90$^{\phi}$ & \multirow{2}{*}{$1\times1\times1$} \\
  & 0.92$^{\psi}$  & 0.63$^{\psi}$  & 1.97$^{\psi}$  & 1.64$^{\psi}$  & 1.37$^{\psi}$  &  1.06$^{\psi}$  &  \\
\bottomrule

\end{tabular}
\end{adjustbox}
\end{table}

\newpage

\section*{Supplementary Note 4. Structural Descriptors}

\subsection*{Sn Off-Centering Magnitude}
\begin{equation}
\Delta_{\mathrm{off}}
=
\left\lVert
\mathbf{r}_{\mathrm{Sn}}
-
\frac{1}{6}\sum_{i=1}^{6}\mathbf{r}_i
\right\rVert
\end{equation}

\subsection*{Off-Centering Orientation Order Parameter}
\begin{equation}
S
=
\left\lVert
\frac{1}{N_{\mathrm{Sn}}}
\sum_{j=1}^{N_{\mathrm{Sn}}}
\frac{\boldsymbol{\delta}_j}{\lVert\boldsymbol{\delta}_j\rVert}
\right\rVert,
\qquad
\boldsymbol{\delta}_j
=
\mathbf{r}_{\mathrm{Sn}}^{(j)}
-
\frac{1}{6}\sum_{i=1}^{6}\mathbf{r}_i^{(j)}
\end{equation}

\subsection*{Sn-I-Sn Bond Angle}

\begin{equation}
\theta
=
\cos^{-1}\!
\left(
\frac{
(\mathbf{r}_{\mathrm{Sn_1}}-\mathbf{r}_{I})
\cdot
(\mathbf{r}_{\mathrm{Sn_2}}-\mathbf{r}_{I})
}{
\lVert\mathbf{r}_{\mathrm{Sn_1}}-\mathbf{r}_{I}\rVert\,
\lVert\mathbf{r}_{\mathrm{Sn_2}}-\mathbf{r}_{I}\rVert
}
\right)
\end{equation}

\subsubsection*{Signed tilt angle.}
\begin{equation}
\tilde{\theta}
=
\mathrm{sgn}
\!\left[
\left( (\mathbf{r}_{\mathrm{Sn_1}}-\mathbf{r}_{I})
\times
(\mathbf{r}_{\mathrm{Sn_2}}-\mathbf{r}_{I}) \right)
\cdot
(\mathbf{r}_{\mathrm{Sn_2}}-\mathbf{r}_{\mathrm{Sn_1}})
\right]
\,(180^\circ - \theta)
\end{equation}

{\small
Here, $\theta$ is the unsigned Sn-I-Sn bond angle, while $\tilde{\theta}$ assigns a sign based on the orientation of the $(\mathrm{Sn_1}, I, \mathrm{Sn_2})$ triad relative to the Sn-Sn axis, distinguishing in-phase and out-of-phase tilts.
}

\subsection*{Octahedral Anisotropy Index}

\begin{equation}
A_{\mathrm{rh}}
=
\frac{
\bar{r}_{\mathrm{long}} - \bar{r}_{\mathrm{short}}
}{
\bar{r}
},
\qquad
\bar{r} = \frac{1}{6}\sum_{i=1}^{6} r_i
\end{equation}

{\small
Here, $\bar{r}_{\mathrm{short}}$ is the mean of the three shortest Sn-I bond lengths,
and $\bar{r}_{\mathrm{long}}$ is the mean of the three longest Sn-I bond lengths.
}

\subsection*{Effective Orbital Overlap Proxy (Normalized)}
\begin{equation}
O_{\mathrm{eff,norm}}
=
\frac{
\dfrac{\cos^{2}\theta}{r_{1}^{2} r_{2}^{2}}
}{
\left\langle
\dfrac{\cos^{2}\theta}{r_{1}^{2} r_{2}^{2}}
\right\rangle_{\mathrm{ref}}
}
\end{equation}
{\small
Here, “ref’’ denotes the undistorted cubic reference structure}
\newpage
\section*{Symbol Legend}

\begin{table}[h!]
\centering
\begin{tabular}{ll}
\hline
Symbol & Meaning \\
\hline
$\mathbf{r}_{\mathrm{Sn}}$ & Cartesian coordinate of a Sn atom \\
$\mathbf{r}_i$ & Coordinates of the six I atoms in a SnI$_6$ octahedron \\
$\mathbf{r}_{\mathrm{Sn_1}}, \mathbf{r}_{\mathrm{Sn_2}}$ & Coordinates of Sn atoms in a Sn-I-Sn bridge \\
$\mathbf{r}_{I}$ & Coordinate of the bridging I atom \\
$r_i$ & Sn-I bond length to the $i$-th I atom \\
$\bar r$ & Mean of the six Sn-I bond lengths \\
$r_1, r_2$ & Sn-I bond lengths in a specific Sn-I-Sn bridge \\
$\boldsymbol{\delta}_j$ & Off-centering vector for the $j$-th Sn atom \\
$N_{\mathrm{Sn}}$ & Number of Sn atoms in the supercell \\
$\theta$ & Unsigned Sn-I-Sn bond angle \\
$\tilde{\theta}$ & Signed Sn-I-Sn tilt angle \\
$A_{\mathrm{rh}}$ & Rhombohedral octahedral anisotropy index \\
$O_{\mathrm{eff,norm}}$ & Normalized effective orbital overlap proxy \\
\hline
\end{tabular}
\end{table}

\newpage

\section*{Supplementary Note 5. Wannier-Center Analysis of Lone-Pair Rotational Dynamics}

For each Sn site, the instantaneous lone-pair orientation vector is defined as
\begin{equation}
    \hat{v}_i(t)
    = \frac{\mathbf{r}_{\mathrm{WC},i}(t) - \mathbf{r}_{\mathrm{Sn},i}(t)}
           {\left\lVert \mathbf{r}_{\mathrm{WC},i}(t) -
                        \mathbf{r}_{\mathrm{Sn},i}(t) \right\rVert},
\end{equation}
where $\mathbf{r}_{\mathrm{Sn},i}(t)$ and $\mathbf{r}_{\mathrm{WC},i}(t)$ are the
positions of the Sn atom and its associated Wannier center at time $t$.

The second-order Legendre rotational time-correlation function is
\begin{equation}
    C_{\mathrm{rot}}(\tau)
    = \Big\langle
        P_2\!\left[\hat{v}(t)\cdot\hat{v}(t+\tau)\right]
      \Big\rangle_t,
\end{equation}
with
\begin{equation}
    P_2(x) = \tfrac{1}{2}\left(3x^2 - 1\right).
\end{equation}

The rotational correlation time is defined by
\begin{equation}
    C_{\mathrm{rot}}(\tau_{\mathrm{rot}}) = e^{-1}.
\end{equation}

The corresponding rotational frequency in wavenumbers (cm$^{-1}$) is
\begin{equation}
    \nu = \frac{1}{\tau_{\mathrm{rot}}\,c},
\end{equation}
where $c = 2.9979\times10^{-5}\,\mathrm{cm/fs}$.

\newpage